# Spontaneous radiation of a two-level atom into multipole modes of a plasmonic nanoparticle

E. S. Andrianov,[1,2] A. A. Pukhov,[1,2,3] A. P. Vinogradov,[1,2,3] A. V. Dorofeenko,[1,2,3] and A. A. Lisyansky[4]

[1]All-Russia Research Institute of Automatics, 22 Sushchevskaya, Moscow 127055, Russia
[2]Moscow Institute of Physics and Technology, Moscow Region, Dolgoprudny, Russia
[3]Institute for Theoretical and Applied Electromagnetics RAS, Moscow, Russia
[4]Department of Physics, Queens College of the City University of New York, Queens, NY 11367, USA

We consider the relaxation of an excited two-level system (TLS) positioned near a spherical plasmonic nanoparticle (NP). The transition frequency of the TLS is assumed to coincide with the frequency of the condensation point of NP plasmonic resonances. We show that the relaxation of the TLS excitation is a two-step process. Following an initial exponential decay, the TLS breaks in to Rabi oscillations. Depending upon the distance between the TLS and NP, the probability of the TLS being in the excited state exhibits either chaotic or nearly regular oscillations. In the latter case, the eigenfrequency of the TLS-NP system coincides with one of NP multipole modes.

## I. INTRODUCTION

The problem of the giant decrease of the radiative relaxation time of atoms near metallic NPs has attracted considerable interest in the last decade due to explosive growth of nanoplasmonics.[1-5] The drop in relaxation time is much more complicated than simple cases of the exponential fall-off of a TLS excitation into a continuum of modes[6] or the Rabi oscillations when the TLS energy is transferred into a single resonant mode.[7] In the case of a metallic NP, the energy can be transferred into an infinite but countable set of modes which spectrum has a condensation point. It has been shown recently[8] that the rate of spontaneous exponential nonradiative decay in a TLS, which is in resonance with the dipole mode of a lossy plasmonic sphere so that its resonant frequency is far from the condensation point, increases by several orders of magnitude thanks to accounting of higher non-resonant multipole modes. This agrees qualitatively with ideas of Ref. 9 that though the contribution of higher multipoles to photon radiation is smaller than that of the dipole, Joule dissipation of higher multipoles is significantly greater than the dissipation of the dipole. The increase of dissipation leads to a decrease of the life-time of the emitter. It is not intuitively clear how a TLS, which is in resonance with the NP condensation point would decay. One might expect that the relaxation of this system due to coupling to an infinite number of modes is similar to the case of coupling to a continuum of modes. On the other hand, the relaxation into the condensation point may resemble the relaxation into a single resonant mode.

In this paper, we study the spontaneous relaxation of a TLS, with a transition frequency in resonance with the frequency of the condensation point of higher multipole resonances. We show that the relaxation of the TLS into a condensation point of the NP plasmonic spectrum is quite unusual. It occurs in two stages. In the initial stage, it has the exponential character due to the existence of infinite albeit countable number of modes to which it couples. This is similar to the case of continuum of modes. However, the system relaxes not into the ground state but towards a quasi-stationary state. Then, in the second stage, the exponential decay transforms into Rabi oscillations. In the latter stage, the probability of the TLS to be in the excited state, exhibits either chaotic or nearly regular oscillations depending upon the distance between the TLS and NP.

The problem of atomic relaxation near a metallic NP requires a quantum description of the field and radiating system.[10-13] The simplest approach using the Fermi Golden Rule does not produce the radiation spectrum unless an additional assumption is made about the Lorentzian line shape, which arises from the interaction of the TLS with continuum of modes.[6] To obtain the decay law from the first principles without additional assumptions we use the Weisskopf-Wigner.[7, 13]

## II. TIME EVOLUTION IN THE LIMITING CASES OF CONTINUUM OF MODES AND SINGLE MODE

The general analysis of a TLS interacting with cavity modes is given in Ref. 6. Let us consider a set $|k\rangle$ of resonator modes and excited $|e\rangle$ and ground $|g\rangle$ states of the TLS. In order to describe the relaxation of the excited state of the TLS, we assume that each resonator mode can interact with a continuum of modes $kk'$ of some reservoir. If the corresponding $kk'$ modes are phonons in metal, the interaction reduces to the Joule losses in the NP. So we may take the Hamiltonian of system in the form

$$\hat{H} = \hbar\omega_{TLS}\hat{\sigma}^\dagger\hat{\sigma} + \hbar\sum_k \omega_k \hat{a}_k^\dagger \hat{a}_k + \hbar\sum_k \gamma_k(\hat{a}_k^\dagger\hat{\sigma} + \hat{\sigma}^\dagger\hat{a}_k) + \hbar\sum_{k'}\omega_{k'}\hat{b}_{k'}^\dagger\hat{b}_{k'} + \hbar\sum_{kk'}\Gamma_{kk'}(\hat{a}_k^\dagger\hat{b}_{kk'} + \hat{b}_{kk'}^\dagger\hat{a}_k). \quad (1)$$

In the Hamiltonian (1), the first term corresponds to the TLS energy ($\omega_{TLS}$ is the TLS transition frequency, $\hat{\sigma}$ is the transition operator between ground and excited states of the TLS), the second term describes the energy of each multipole ($\omega_k$ is the resonance frequency of the $k$-th multipole mode of the NP, $\hat{a}_k$ is the plasmon annihilation operator), the third term corresponds to the interaction of the TLS and the $k$-th multipole mode ($\gamma_k$ is the interaction constant), the fourth term corresponds to the energy of the thermal reservoir ($\omega_{k'}$ is the eigenfrequency of $k'$-th multipole mode of the NP, $\hat{b}_{k'}$ is the annihilation operator of this mode), and the last term describes the interaction between $k$-th and $k'$-th multipole modes of the thermal bath ($\Gamma_{kk'}$ is the interaction constant).

Let us expand the wave function of the system "atom + resonator modes + reservoir modes" over the stationary states in absence of interactions

$$\Psi(t) = A(t)\exp(-i\omega_{TLS}t)|e,0,0\rangle + \sum_k B_k(t)\exp(-i\omega_k t)|g,1_k,0\rangle$$
$$+ \sum_{k,k'} C_{kk'}(t)\exp(-i\omega_{kk'}t)|g,0,1_{kk'}\rangle, \quad (2)$$

where $|e,0,0\rangle$ denotes the state in which only the atom is excited, $|g,1_k,0\rangle$ is the state in which only the $k$-th mode of the resonator is excited, and $|g,0,1_{kk'}\rangle$ denotes the state in which only $k'$-th wall mode of the $k$-th mode of the resonator is excited.

Using the Schrödinger equation with Hamiltonian (1) and expansion (2) we obtain the system of equations

$$i\dot{A}(t) = \sum_k \gamma_k^* B_k(t)\exp(-i(\omega_k - \omega_{TLS})t), \quad (3)$$

$$i\dot{B}_k(t) = \gamma_k A(t)\exp(i(\omega_k - \omega_{TLS})t) + \sum_{k'} \Gamma_{kk'}^* C_{kk'}(t)\exp(-i(\omega_{kk'} - \omega_k)t), \quad (4)$$

$$i\dot{C}_{kk'}(t) = \Gamma_{kk'} B_k(t)\exp(i(\omega_{kk'} - \omega_k)t). \quad (5)$$

Taking into account the initial conditions $A(0)=1$, $B_k(0)=0$, $C_{kk'}(0)=0$ and using the Fourier transformation

$$\alpha(q) = \frac{1}{2\pi}\int_0^\infty A(t)\exp(iqt)dt, \quad (6)$$

$$\beta_k(q) = \frac{1}{2\pi}\int_0^\infty B_k(t)\exp(-i(\omega_k - \omega_{TLS})t)\exp(iqt)dt, \quad (7)$$

$$S_{kk'}(q) = \frac{1}{2\pi}\int_0^\infty C_{kk'}(t)\exp(-i(\omega_{kk'} - \omega_{TLS})t)\exp(iqt)dt, \quad (8)$$

we obtain an algebraic system of equations

$$-\frac{i}{2\pi} + q\alpha(q) = \sum_k \gamma_k^* \beta_k(q), \quad (9)$$

$$(q-\Delta_k)\beta_k = \gamma_k \alpha(q) + \sum_{k'} \Gamma_{kk'}^* S_{kk'}(q), \quad (10)$$

$$(q-(\Delta_{kk'}+\Delta_k))S_{kk'}(q) = \Gamma_{kk'}\beta_k, \quad (11)$$

where $\Delta_k = \omega_k - \omega_{TLS}$ and $\Delta_{kk'} = \omega_{kk'} - \omega_{TLS}$.

The solutions to Eqs. (10) and (11) are

$$S_{kk'}(q) = \frac{\Gamma_{kk'}\beta_k}{q-(\Delta_k+\Delta_{kk'})}, \tag{12}$$

$$\beta_k(q) = \frac{\gamma_k \alpha(q)}{q-\Delta_k - \sum_{k'}\frac{|\Gamma_{kk'}|^2}{q-(\Delta_k+\Delta_{kk'})+i0}}. \tag{13}$$

In our case, the resonator modes are all multipole plasmonic modes. They are collective excitations of electrons interacting with boson modes of the thermal reservoir (i.e., with phonons). This interaction represents Joule losses in metal. In Eq. (13), the interaction between modes and the reservoir is described by the sum in the denominator which can be expressed as

$$\sum_{k'}\frac{|\Gamma_{kk'}|^2}{q-(\Delta_k+\Delta_{kk'})} = \int dk' \rho(k') \frac{|\Gamma_{kk'}|^2}{q-(\Delta_k+\Delta_{kk'})}, \tag{14}$$

where $\rho(k')$ is the density of states of the thermal bath. Using the standard regularization procedure, $(x+i0)^{-1} = -i\pi\delta(x) + P(1/x)$, and neglecting the Lamb shift we obtain the effective relaxation rate $g_k$, which is equal to the linewidth of the $k$-th mode

$$\int dk' \rho(k') \frac{|\Gamma_{kk'}|^2}{q-(\Delta_k+\Delta_{kk'})} \approx -i\pi\rho(k')|\Gamma_{kk'}|^2\bigg|_{\Delta_{kk'}=q-\Delta_k} = -ig_k. \tag{15}$$

Thus, the thermal bath with the continuum of bosonic modes induces the exponential decay to the $k$-th mode of the system which interacts with this bath.[4, 7]

Using Eq. (15) we obtain the following solution for Eqs. (9)-(11):

$$\alpha(q) = \frac{i}{2\pi} \frac{1}{q - \sum_k \frac{|\gamma_k|^2}{q-\Delta_k+ig_k}}, \tag{16}$$

$$\beta_k(q) = \frac{\alpha(q)\gamma_k}{q-\Delta_k+ig_k}. \tag{17}$$

Substituting Eqs. (16) and (17) into Eqs. (6) and (7) and using the inverse Fourier transform we obtain the probability amplitudes

$$A(t) = \frac{1}{2\pi i} \int_{i\delta+\infty}^{i\delta-\infty} \frac{\exp(-iqt)\,dq}{q - \sum_k \frac{|\gamma_k|^2}{q - \Delta_k + ig_k}}, \tag{18}$$

$$B_k(t) = \frac{1}{2\pi i} \int_{i\delta+\infty}^{i\delta-\infty} \frac{\gamma_k \exp(-iqt)\exp(-it/k)\,dq}{(q - \Delta_k + ig_k)\left(q - \sum_k \frac{|\gamma_k|^2}{q - \Delta_k + ig_k}\right)}. \tag{19}$$

The value of $\delta$ in Eqs. (18)-(19) should be greater than real parts of any pole of the integrands in these equations. Then, all the poles are inside of the integration contour, which is the straight line $(i\delta + \infty, i\delta - \infty)$ closed into the lower half-plane.

Form Eqs. (18)-(19) we can obtain the main results concerning atom dynamics. As an example, we consider two limiting cases. In the first one, we obtain the relaxation of a two-level atom in free space ($|\gamma_k|^2 = \omega_k \mu_{TLS}^2 \cos^2\theta / 2\varepsilon_0 \hbar V$) with continuum of modes.[4, 6] To begin with, one can use the standard summation procedure on free space modes,

$$\sum_{\mathbf{k}} \to 2\frac{V}{(2\pi)^3} \int_0^{2\pi} d\varphi \int_0^{\pi} d\theta \sin\theta \int_0^{\infty} k^2 dk, \tag{20}$$

and then use the regularization procedure described above to obtain:

$$\sum_k \frac{|\gamma_k|^2}{q - \Delta_k} \to \frac{\mu_{TLS}^2}{6\pi^2 \varepsilon_0 \hbar c^3} \int_0^{\infty} \frac{\omega^3 d\omega}{q - (\omega - \omega_{TLS})}$$
$$\approx -i \frac{\mu_{TLS}^2 \omega_{TLS}^3}{6\pi \varepsilon_0 \hbar c^3} + \frac{\mu_{TLS}^2}{6\pi^2 \varepsilon_0 \hbar c^3} P\int_0^{\infty} \frac{\omega^3 d\omega}{\omega - \omega_{TLS}} = -i\gamma/2 + \Delta\omega_L, \tag{21}$$

where we $\gamma$ and $\Delta\omega_L$ represent the dissipation rate and the Lamb shift, respectively. Now, Eq. (18) takes the form

$$A(t) = \frac{1}{2\pi i} \int_{i\delta+\infty}^{i\delta-\infty} \frac{\exp(-iqt)\,dq}{q - \Delta\omega + i\gamma/2} = \exp(-(\gamma/2 + i\Delta\omega)t), \tag{22}$$

which corresponds to the exponential decay to free space modes.

The second case corresponds to decay of a TLS in a single mode cavity.[7] In this approximation, $|\gamma_k|^2 = \omega_R^2$, $g_k = 0$ and we can find poles of the integrand of Eq. (18): $q = \pm\omega_R$ which corresponds to oscillation with the Rabi frequency:

$$A(t) = \cos(\omega_R t). \tag{23}$$

Thus, as expected, the interaction with system with continuous number of modes results in the exponential decay while the interaction with single mode results in the Rabi oscillations and absence of decay.

## III. MULTIPOLE MODES OF THE NANOPARTICLE

Now, we consider the relaxation of the TLS interacting with the countable number of modes whose spectrum has a condensation point. It is hard to predict the behavior of this system qualitatively because, on the one hand, we have an infinite number of modes that should cause an exponential decay, on the other hand, the set of modes is discrete and one can expect that the interaction with each mode may lead to the Rabi oscillations. As an example we consider a spherical plasmonic NP with the dielectric permittivity $\varepsilon = -1$.

In this and the next subsections, we discuss the mode structure of a plasmon spherical NP and perform its quantization. These modes are determined by eigensolutions of the Laplace equation for the scalar potential $\varphi$

$$\nabla(\varepsilon(\mathbf{r})\nabla\varphi(\mathbf{r})) = 0 \tag{24}$$

with the boundary conditions on the sphere surface

$$\varepsilon_{in} \left.\frac{\partial \varphi^{in}}{\partial \mathbf{n}}\right|_{|\mathbf{r}|=a} = \varepsilon_{out} \left.\frac{\partial \varphi^{out}}{\partial \mathbf{n}}\right|_{|\mathbf{r}|=a}, \tag{25}$$

where $a$ is the radius of the sphere, $\varepsilon_{in}$, $\varphi_{in}$ and $\varepsilon_{out}$, $\varphi_{out}$ are dielectric permittivities and scalar potentials inside and outside of the sphere, respectively. For the sake of simplicity, we assume that $\varepsilon_{out} = 1$. Then the solution of Eq. (24) with boundary condition (25) has the form[5]

$$\varphi_{nm} = \begin{cases} (r/a)^n Y_{nm}(\theta,\varphi), & r \leq a \\ (r/a)^{n+1} Y_{nm}(\theta,\varphi), & r > a \end{cases}, \tag{26}$$

where $Y_{nm}(\theta,\varphi)$ are spherical functions. The $n$-th frequency is determined by the condition

$$\varepsilon_{in}(\omega_n) = -\frac{n+1}{n}. \tag{27}$$

Below, we assume that the dispersion of the metal NP is described by the Drude formula

$$\varepsilon_{in}(\omega) = 1 - \frac{\omega_{pl}^2}{\omega^2}, \tag{28}$$

where $\omega_{pl}$ is the plasmon frequency of metal. Eqs. (27) and (28) determine resonance frequencies

$$\omega_n = \omega_{pl}\sqrt{\frac{n}{2n+1}}. \tag{29}$$

Note that eigenfunctions determined by Eq. (26) are degenerated: $2n + 1$ eigenfunctions corresponds to the *n*-th resonant frequency as parameter *m* runs from $-n$ to $n$. When $n \to \infty$, the resonant frequency $\omega_n \to \omega_{pl}/\sqrt{2}$, which is the condensation point of eigenfrequencies.

## A. Quantization

Before we start quantizing electromagnetic field of the spherical NP, we have to note that since the Laplace equation only describes the static (near) field of the NP, in this approximation, the electric field does not change in time and the magnetic field is equal to zero. Because quantization of the electromagnetic field and the concept of the excitation quant are related to the part of the vector potential which depends on time, in this approximation, the quantization seems impossible. This can be resolved if one considers the *quasistatic* approximation in which the dependence of the potential on coordinates is given by Eq. (26), while the dependence on time is simply harmonic oscillation with the frequency of the corresponding resonance. That is, the electric field has the form $\mathbf{E}_{nm}(\mathbf{r},t) = -\nabla \varphi_{nm}(\mathbf{r})\exp(i\omega_n t)$. Then the electric field is a harmonic oscillator that can be quantized in a standard way (see, e.g., Ref. 11).

At the frequency of the *n*-th plasmon resonance, the corresponding mode of the electric field of the NP obeys the harmonic oscillator equation with the frequency of the plasmonic resonance

$$\ddot{\mathbf{E}}_{nm} + \omega_n^2 \mathbf{E}_{nm} = 0. \tag{30}$$

Now, we can introduce creation, $\hat{\tilde{a}}_{nm}^\dagger(t)$, and annihilation, $\hat{\tilde{a}}_{nm}(t)$, Bose-operators of the dipole moment of the surface plasmon excited at the NP. These operators satisfy the commutation relation $\left[\hat{\tilde{a}}_{nm}(t),\hat{\tilde{a}}_{nm}^\dagger(t)\right]=1$. The operator of the electric field can be expressed as

$$\hat{\mathbf{E}}_{nm} = -\tilde{E}_{nm}\nabla\varphi_{nm}\left(\hat{\tilde{a}}_{nm} + \hat{\tilde{a}}_{nm}^+\right)/\sqrt{2}, \tag{31}$$

where the dimensional factor $\tilde{E}_{nm}$ is the "quant" of the electric field. The Hamiltonian of the harmonic oscillator has the form $\hat{H}_{nm} = \hbar\omega_n\left(\hat{\tilde{a}}_{nm}^+\hat{\tilde{a}}_{nm} + 1/2\right)$.

In order to obtain $\tilde{E}_{nm}$, one can compare the energy of one quant with the energy of *n*-th plasmon mode:

$$\hbar\omega_n = \frac{1}{8\pi}\int_V \left.\frac{\partial(\omega\,\mathrm{Re}\,\varepsilon(\omega))}{\partial\omega}\right|_{\omega_n} |\tilde{E}_{nm}|^2 \nabla\varphi_{nm}\cdot\nabla\varphi_{nm}^* dV. \qquad (32)$$

Integral (32) can be expressed as

$$\frac{1}{8\pi}\int_V \left.\frac{\partial(\omega\,\mathrm{Re}\,\varepsilon(\omega))}{\partial\omega}\right|_{\omega_n} |\tilde{E}_{nm}|^2 \nabla\varphi_{nm}\cdot\nabla\varphi_{nm}^* dV$$

$$= \frac{|\tilde{E}_{nm}|^2}{8\pi}\left(\int_V \omega\left.\frac{\partial(\mathrm{Re}\,\varepsilon(\omega))}{\partial\omega}\right|_{\omega_n} \nabla\varphi_{nm}\cdot\nabla\varphi_{nm}^* dV + \int_V \mathrm{Re}\,\varepsilon(\omega)\big|_{\omega_n} \nabla\varphi_{nm}\cdot\nabla\varphi_{nm}^* dV\right). \qquad (33)$$

Since the scalar potential satisfies the Laplace equation, $\nabla(\mathrm{Re}\,\varepsilon(\omega_n)\nabla\varphi)=0$, and the boundary conditions (25), the second term in the right hand part of Eq. (33) is equal to zero:

$$\int_V \mathrm{Re}\,\varepsilon(\omega)\big|_{\omega_n}\nabla\varphi_{nm}\cdot\nabla\varphi_{nm}^* dV$$

$$= \left(\int_{V_{in}} dV_{in} + \int_{V_{out}} dV_{out}\right)\varphi_{nm}^*\nabla(\mathrm{Re}\,\varepsilon(\omega)\nabla\varphi_{nm})\big|_{\omega_n} + \int_{\partial V}\varphi_{nm}^*\left(\varepsilon_{in}\frac{\partial\varphi_{nm}^{in}}{\partial \mathbf{r}} - \varepsilon_{out}\frac{\partial\varphi_{nm}^{out}}{\partial \mathbf{r}}\right)\cdot\mathbf{n}\, dA = 0, \qquad (34)$$

where the second integral in the right hand part of Eq. (34) is calculated over the surface of the NP. We assume that the medium outside the NP is dispersionless, $\varepsilon_{out}=1$, therefore, $\partial\varepsilon_{out}/\partial\omega=0$. Substituting $\varphi_{nm}$ defined by Eq. (26) into the first term in the right hand part of Eq. (34) and assuming that inside the NP, the dielectric permittivity does not depend on coordinates and is described by Eq. (27) we obtain

$$\frac{1}{8\pi}\int_{V_{in}}\omega\left.\frac{\partial\,\mathrm{Re}\,\varepsilon(\omega)}{\partial\omega}\right|_{\omega_n}\nabla\varphi_{nm}\cdot\nabla\varphi_{nm}^* dV = \frac{\omega_{pl}^2}{4\pi\omega_n^2}\int_{V_{in}}\nabla\varphi_{nm}^*\nabla\varphi_{nm}dV$$

$$= \frac{\omega_{pl}^2}{4\pi\omega_n^2}\left(\int_{\partial V_{in}}\varphi_{nm}^*\frac{\partial\varphi_{nm}}{\partial \mathbf{n}}dS - \int_{V_{in}}\varphi_{nm}^*\Delta\varphi_{nm}dV\right) = \frac{\omega_{pl}^2}{4\pi\omega_n^2}\int_\Omega na Y_{nm}^* Y_{nm}d\Omega = \frac{2n+1}{4\pi}a. \qquad (35)$$

When obtaining Eq. (35) we use the normalization of the spherical functions, $\int_\Omega Y_{nm}^* Y_{nm} d\Omega = 1$, and Eq. (29). Finally, we arrive at

$$\hbar\omega_{SP} = |\tilde{E}_{nm}|^2 a(2n+1)/4\pi, \qquad (36)$$

so that $\tilde{E}_{nm} = \sqrt{4\pi\hbar\omega_{nm}/a(2n+1)} = E_{nm}\sqrt{2}$. Thus, the operator of the electric field of the *n*-th mode of the NP is

$$\hat{\mathbf{E}}_{nm} = -E_{nm}\nabla\varphi_{nm}\left(\hat{\tilde{a}}_{nm} + \hat{\tilde{a}}^{\dagger}_{nm}\right). \tag{37}$$

## B. NP-TLS Interaction

The Hamiltonian of a TLS can be represented as

$$\hat{H}_{TLS} = \hbar\omega_{TLS}\hat{\tilde{\sigma}}^{\dagger}\hat{\tilde{\sigma}}, \tag{38}$$

where $\hat{\tilde{\sigma}} = |g\rangle\langle e|$ is the operator of the transition between the ground $|g\rangle$ and excited $|e\rangle$ states of the TLS. We assume that the atom has a dipole moment only, which is equal to $\hat{\mathbf{d}}_{TLS} = \mu_{TLS}\mathbf{e}_{TLS}\left(\hat{\tilde{\sigma}}(t) + \hat{\tilde{\sigma}}^{\dagger}(t)\right)$, where $\mu_{TLS} = \langle e|er|g\rangle$ is the value of the dipole transition and $\mathbf{e}_{TLS}$ is the unit vector in the direction of the TLS dipole moment. When the TLS transition frequency $\omega_{TLS}$ is in resonance with the frequency of the condensation point, it is convenient to represent operators $\hat{\tilde{a}}(t)$ and $\hat{\tilde{\sigma}}(t)$ as $\hat{\tilde{a}}(t) = \hat{a}(t)e^{-i\omega_{TLS}t}$ and $\hat{\tilde{\sigma}}(t) \equiv \hat{\sigma}(t)e^{-i\omega_{TLS}t}$, where $\hat{a}(t)$ and $\hat{\sigma}(t)$ are slowly varying operators. Now, we apply the rotating wave approximation[7] in which fast oscillating terms proportional to $e^{\pm 2i\omega_{TLS}t}$ are neglected. In this approximation, the interaction $\hat{V} = -\hat{\mathbf{d}}\cdot\hat{\mathbf{E}}$ of the TLS dipole moment with the NP field, $\hat{\mathbf{E}} = -\sum_{nm}E_{nm}\nabla\varphi_{nm}\left(\hat{\tilde{a}}_{nm} + \hat{\tilde{a}}^{+}_{nm}\right)$, has the form[7]

$$\hat{V} = \hbar\sum_{nm}\gamma_{nm}\left(\hat{a}^{\dagger}_{nm}\hat{\sigma} + \hat{\sigma}^{\dagger}\hat{a}_{nm}\right), \tag{39}$$

where the Rabi interaction constant of the TLS dipole moment with the *nm*-th mode of the NP field, $\gamma_{nm}$, has to be determined,.

In the problem under consideration, there are two characteristic vectors: the vector of the direction of the dipole moment of the atom, which we assume is determined by the atom's structure, and the vector directed from the center of the atom to the center of the NP. Below, for the sake of simplicity, we assume that these vectors are collinear, i.e. the dipole is positioned "over" the NP. The convenient reference frame for this geometry is such that the z-axis passes through the centers of the atom and the NP, so that $\theta = 0$. In this reference frame, the atom's dipole interacts with radial component of the field only. In addition, the atom only interacts with symmetric configurations of the NP field for which $m = 0$. The latter can be shown by using the representation of the spherical functions via Legendre polynomials:

$$Y_{nm}(\theta,\varphi) = \sqrt{\frac{2n+1}{4\pi}\frac{(n-m)!}{(n+m)!}}P_n^m(\cos\theta)\exp(im\varphi). \tag{40}$$

Since $\theta = 0$, we have $P_n^0(1) = 1$ and $P_n^m(\cos\theta) = \sin^m\theta \dfrac{d^m}{d(\cos\theta)^m} P_n^0(\cos\theta) = 0$, so that only the harmonic with $m = 0$ survives. Then, because $E_{nm\ r}(\theta,\varphi) \propto Y_{nm}(\theta,\varphi)$, only $E_{n,m=0}$ should be taken into account. This allows us to represent the interaction constant in the form

$$|\gamma_n|^2 = \left|\frac{E_{n0}\nabla\varphi_{n0}\mu_{TLS}}{\hbar}\right|^2 = \frac{2\pi\omega_n}{\hbar a(2n+1)}\frac{(n+1)^2}{a^2}\left(\frac{a}{r}\right)^{2(n+2)}\frac{2n+1}{4\pi}|\mu_{TLS}|^2$$
$$= \frac{|\mu_{TLS}|^2 \omega_{pl}}{2\hbar a^3}\xi^{2(n+2)}(n+1)^2 n^{1/2}(2n+1)^{-1/2},$$
(41)

where $\xi = a/r_0$, $a$ is the radius of NP and $r_0$ is the distance between the NP and the TLS.

The full Hamiltonian of the TLS interacting with the NP has the form

$$\hat{H} = \hbar\sum_n \omega_n \hat{a}_n^\dagger \hat{a}_n + \hbar\omega_{TLS}\hat{\sigma}^\dagger\hat{\sigma} + \hbar\sum_n \gamma_n(\hat{a}_n^\dagger\hat{\sigma} + \hat{\sigma}^\dagger\hat{a}_n),$$
(42)

where $\omega_n$ and $\gamma_n$ are defined by Eqs. (29) and (41), respectively.

## IV. TIME EVOLUTION OF THE TWO-LEVEL SYSTEM NEAR THE CONDENSATION POINT OF EIGENMODES OF NANOPARTICLE

### A. Lossless system

Now, we consider the problem of the excitation of plasmonic modes having a condensation point. In this case, $\Delta_k = \omega_k - \omega_{TLS}$ and $\Delta_{kk'} = \omega_{kk'} - \omega_{TLS}$. Since $\Delta_k = \omega_{pl}\sqrt{k/(2k+1)} - \omega_{pl}/\sqrt{2} \approx -\omega_{pl}/(4\sqrt{2}k)$, we express all the frequencies in the units of $\omega_0 = \omega_{pl}/4\sqrt{2}$, so that

$$\Delta_k = -\omega_0/k,\quad |\gamma_k|^2 = \gamma^2\xi^{2k}(k+1)^2 k^{1/2}(2k+1)^{-1/2},$$
(43)

where $\gamma^2 = |\mu_{TLS}|^2 \omega_{pl}\xi^4/2\hbar a^3 = 16|\mu_{TLS}|^2 \xi^4\omega_0^2/(\omega_{pl}\hbar a^3)$. Using the explicit dependence of $\gamma_k$ and $\Delta_k$ on the mode number $k$, Eq. (31), we obtain

$$A(t) = \frac{1}{2\pi i}\int_{i\varepsilon+\infty}^{i\varepsilon-\infty}\frac{\exp(-iqt)dq}{q - \gamma^2\sum_k \dfrac{\xi^{2k}(k+1)^2 k^{1/2}(2k+1)^{-1/2}}{q + 1/k + ig_k}},$$
(44)

$$B_k(t) = \frac{1}{2\pi i} \int_{i\varepsilon+\infty}^{i\varepsilon-\infty} \frac{\gamma \xi^k (k+1) k^{1/4} (2k+1)^{-1/4} \exp(-iqt)\exp(-it/k)\, dq}{(q+1/k+ig)\left(q - \gamma^2 \sum_k \dfrac{\xi^{2k}(k+1)^2 k^{1/2}(2k+1)^{-1/2}}{q+1/k+ig_k}\right)}. \tag{45}$$

In the next section, we use the obtained equations to investigate the atomic relaxation into multipole modes of a plasmonic NP.

Let us, first, consider the time dynamics of probability amplitudes when attenuation is neglected ($g = 0$). This reflects the physics of the system for small time intervals in which attenuation is not important.

From Eqs. (44) and (45) one can see that the controlling dimensionless parameter of the problem is the ratio of the radius $a$ of the spherical NP to the distance $r_0$ between the two-level atom: $\xi = a/r_0$. In order to study the dynamics of probabilities of excitations of the TLS, $|A(t)|^2$, and the $k$-th mode of resonator $|B_k(t)|^2$, we solve Eqs. (22)-(24) numerically for $g = 0$ for fixed values of $\xi$. The results of the numerical calculations are shown in Fig. 1.

The numerical analysis shows that along with the exponential decay there are irregular Rabi oscillations with the characteristic frequency $\Omega_R$ (see Fig. 1). For the times smaller than the Rabi period, $\gamma^{-1}$, the probability $|A(t)|^2$ falls off exponentially until it reaches a quasi-stationary value, $F(\infty) = \lim_{T\to\infty} \dfrac{1}{T} \int_0^T |A(t)|^2 dt$; after that the oscillations begin. When obtaining these results we assumed that the decay time of the free TLS is much larger than the time interval under consideration. Thus, the exponential decay may be associated with excitation of infinite number of plasmonic modes.

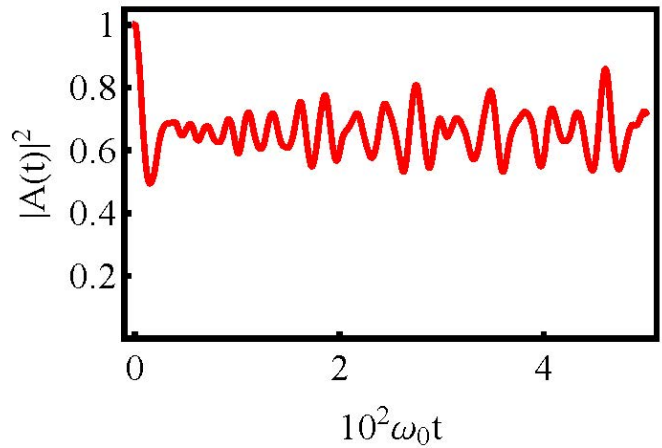

FIG. 1. The time dependence of the probability, $|A(t)|^2$, of the two-level atoms to be in the excited state for $\xi = 0.8$.

Let us analyze the behavior shown in Fig. 1 analytically. For this we estimate the Fourier-transform of the probability amplitude $\alpha(q)$ for the population of the excited state of the TLS, defined by Eqs. (6) and (44). Using an estimate $|\gamma_n|^2 = \gamma^2 \xi^{2n}(n+1)^2 n^{1/2}(2n+1)^{-1/2} \sim \gamma^2 n^2 \xi^{2n}$ for large $n$, $\alpha(q) = \frac{1}{2\pi}\int_0^\infty A(t)\exp(iqt)dt,$ can be expressed as

$$\alpha(q) = \frac{i}{2\pi} \frac{1}{q - \sum_k \frac{|\gamma_k|^2}{q - \Delta_k + ig_k}} \sim \frac{1}{q} \frac{1}{1 - \frac{\gamma^2}{q}\sum_{k=1}^\infty \frac{\xi^{2k}k^2}{q+1/k}}. \tag{46}$$

The asymptotic expansion of the Fourier transform near an infinitely remote point gives the asymptote of the original function near zero. Assuming that in the denominator of Eq. (46) $q \gg 1/k$, we can qualitatively understand the dynamics of the process at initial time $t \ll k$. We have

$$\alpha(q) \sim \frac{1}{q}\frac{1}{1 - \frac{\gamma^2}{q}\sum_{k=1}^\infty \frac{\xi^{2k}k^2}{q+1/k}} = \frac{1}{q}\frac{1}{1 - \frac{\gamma^2}{q}\sum_{k=1}^\infty \frac{\xi^{2k}k^2}{q}\sum_{l=0}^\infty \frac{(-1)^l}{k^l q^l}}$$
$$= \frac{1}{q}\frac{1}{1 - \frac{\gamma^2}{q}\sum_{l=0}^\infty \frac{(-1)^l}{q^{l+1}}\sum_{k=1}^\infty \frac{(\xi^2)^k}{k^{l-2}}} = \frac{1}{q}\frac{1}{1 - \frac{\gamma^2}{q}\sum_{l=0}^\infty \frac{(-1)^l \operatorname{Li}_{l-2}(\xi^2)}{q^{l+1}}}, \tag{47}$$

where $\operatorname{Li}_l(x) = \sum_{k=1}^\infty x^k/l^k$ is polylogarithmic function. Since we are interested in the case when the TLS is near the NP, we assume that $\xi \sim 1$. Then, the main term in the denominator of expression (47) is the polylogarithm with $l = 0$, while the terms with $l = 1,2,...$ produce corrections. Then Eq. (47) can be evaluated as

$$\alpha(q) \sim \frac{1}{q - \frac{\gamma^2}{q}\operatorname{Li}_{-2}(\xi^2) + \frac{\gamma^2}{q^2}\operatorname{Li}_{-1}(\xi^2)} = \frac{1}{q - \frac{\gamma^2}{q}\frac{\xi^2(1+\xi^2)}{(1-\xi^2)^3} + \frac{\gamma^2}{q^2}\frac{\xi^2}{(1-\xi^2)^2}}.$$

The pole of $\alpha(q)$ is equal to the root $q = (\Omega_R + i\Gamma)/2$ of the equation

$$q - \frac{\gamma^2}{q}\frac{\xi^2(1+\xi^2)}{(1-\xi^2)^3} + \frac{\gamma^2}{q^2}\frac{\xi^2}{(1-\xi^2)^2} = 0. \tag{48}$$

Thus, the imaginary part of the pole characterizes the rate of the exponential decay of $A(t)$. Fig. 2 shows the rate of the exponential transition to the quasi-stationary value as a function of $\xi$.

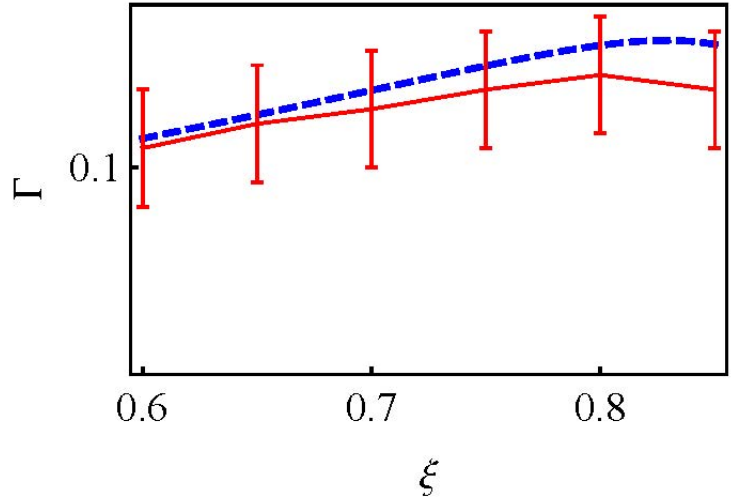

FIG. 2. The dependence of the exponential decay rate, $\Gamma$, on the $\xi$. The results of numerical simulations and analytical evaluation are shown by solid red and dashed blue lines, respectively.

In order to retrieve the characteristic relaxation time from computer simulation data, we consider the function $P(t) = |A(t)|^2 - F(\infty)$ (shown in Fig. 3) and assume that at short times $P(t) = \left(|A(0)|^2 - F(\infty)\right)\exp(-\Gamma t)\cos(\Omega_R t)$. The characteristic parameters, $\Gamma$ and $\Omega_R$, can found by minimizing the quantity $\left|\int_0^T \left(|A(0)|^2 - F(\infty)\right)\exp(-\Gamma t)\cos(\Omega_R t) - \left(|A(t)|^2 - F(\infty)\right) dt\right|$.

As one can see from Fig. 3, the exponential decay modulated by cosine provides a fairly good fit to numerical calculations.

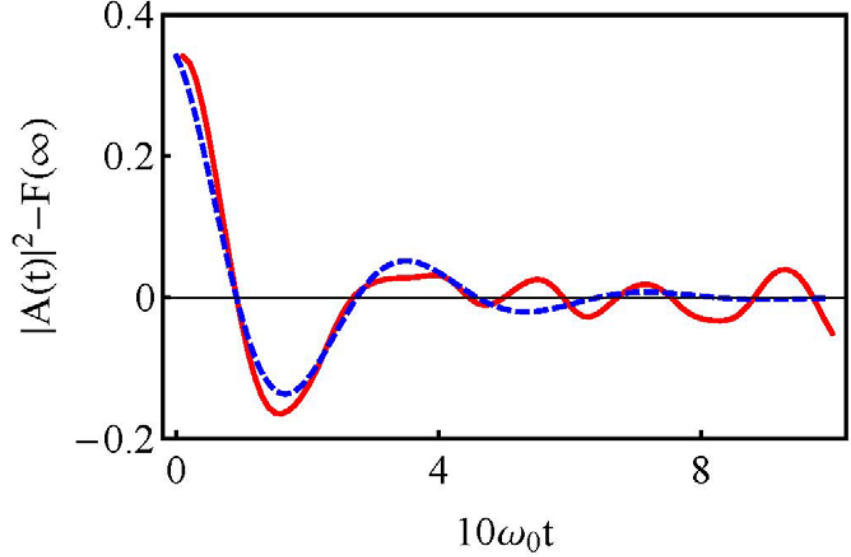

FIG. 3. The dependence of $|A(t)|^2 - F(\infty)$ on time (red solid line) and the approximation function $\exp(-\Gamma t)\cos(\Omega_R t)$ with optimal parameters (blue dashed line) for $\xi = 0.8$.

It turns out that the rate at which the system relaxes to the quasi-stationary state, the parameter $\Gamma$ in the probe function, practically does not depend on $\xi$ (Fig. 2). This can be understood if one takes into account that the exponential character of the decay is usually related to the presence of an infinite number of modes. Since the density of states of multipole modes does not depend on $\xi$, one can expect a weak dependence of $\Gamma$ on $\xi$ as well. Unlike $\Gamma$, the quasi-stationary state $F(\infty)$ does depend on $\xi$ (see Fig. 4).

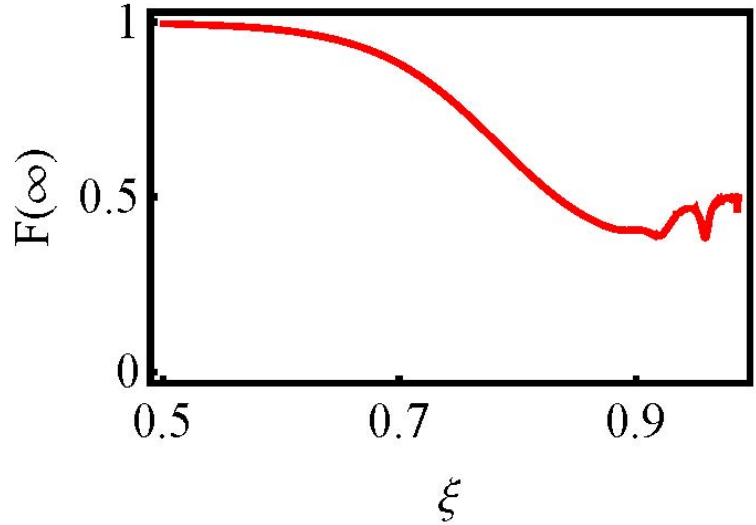

FIG. 4. The dependence of the average probability $F(\infty)$ of a two-level atom to be in the excited state on $\xi$.

For an infinitely large distance between the NP and the TLS, $\xi = 0$, we arrive at the free excited TLS, which should emit a photon and decays exponentially.[6] As we point out above, the characteristic time of this process, $\tau_{free}$, is much larger than the time intervals that we consider. For finite $\xi$, during the characteristic time $\tau \ll \tau_{free}$, the TLS relaxes to a quasi-stationary state, $F(\infty)$, which decreases when $\xi$ grows. Concurrently, due to an increase of coupling constants the frequency of the oscillations (an effective Rabi frequency) increases and the amplitudes of these oscillations grow reaching unity.

For $\xi \geq 1$, the characteristic decay constant $\Gamma$ becomes smaller than $\Omega_R$, and since the averaging obliterates information of the system evolution, our approach ceases to work. Now, to study the system oscillations we should directly investigate $|A(t)|^2$. These oscillations are related to the TLS interaction with an infinite number of higher multipole modes. Our calculations show that for certain distances between the TLS and NP, the probability $|A(t)|^2$ can exhibit almost harmonic oscillations with fixed frequencies. For these distances, peculiarities of the dependence of $\overline{|A(t)|^2}$ on $\xi$ occur. For $\xi > 0.9$, such peculiarities, the nature of which is discussed below, can be seen in Fig. 4. The oscillations corresponding to the peculiarity at $\xi = 0.93$ are shown in Fig. 5.

Since the eigenfrequencies of non-interacting NP and TLS coincide, an increase in the coupling constants leads to the splitting of the eigenfrequency of the joint system. Thus, the eigenfrequencies of the joint system move away from the point of condensation. Consequently, the latter eigenfrequencies get into the range where the joint system feels the discrete character of the plasmonic spectrum. If the eigenfrequency coincides with one of multipole resonance we observe almost regular Rabi oscillations. At such values of $\xi$ the amplitude of the corresponding multipole oscillation predominates the ones of the other multipoles (see Fig. 9).

We can estimate the number of efficiently interacting modes. When $\xi$ is small, the interaction constant depends on the mode number as $\gamma_n \sim n\xi^n$. The mode interacting the most has the number $n_{max} \sim -1/\log\xi$ which corresponds to the interaction constant $\gamma_{max} \sim -1/e\log\xi$. The width of the peak is $\Delta n_{max} \sim -1/\log\xi$. Thus, when the TLS approaches the NP, the peak and its width grow and the peak position is shifted to higher $n$; as $\xi \to 1$, $\gamma_{max} \to \infty$ and $n_{max} \to \infty$ (see Fig. 6). This means that the TLS interacts with of higher multipole modes and the number interacting mode increases (see Fig. 7). This is confirmed by numerical calculations. In Fig. 8, in which average probabilities of excitations of the resonator $k$-th mode, $\overline{|B_k(t)|^2}$, for $k$ =1, 2, 3, and 6 are shown.

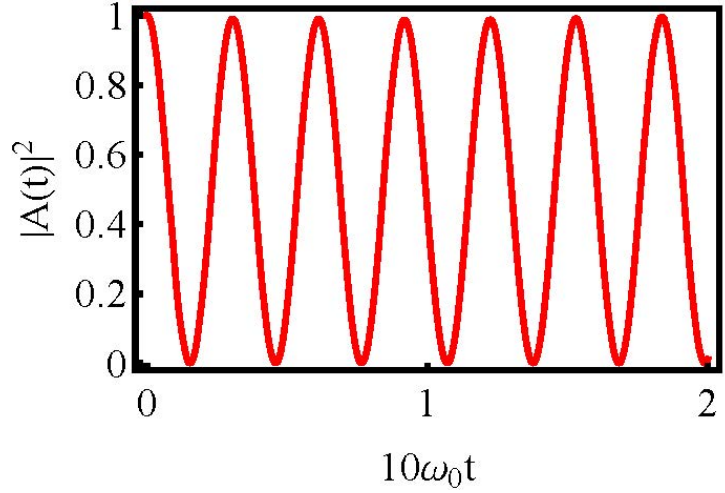

FIG. 5. The dependence of the probability of the two-level atoms to be in the excited state, $|A(t)|^2$, on time for $\xi = 0.93$.

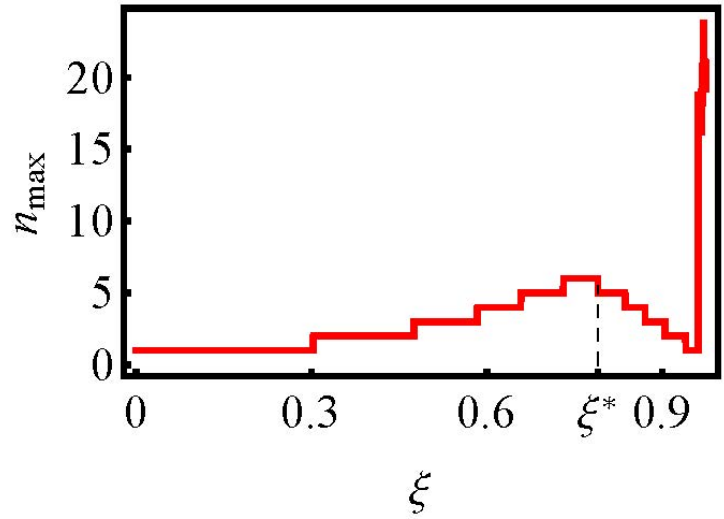

FIG. 6. The dependence of the mode number with maximum energy on $\xi$.

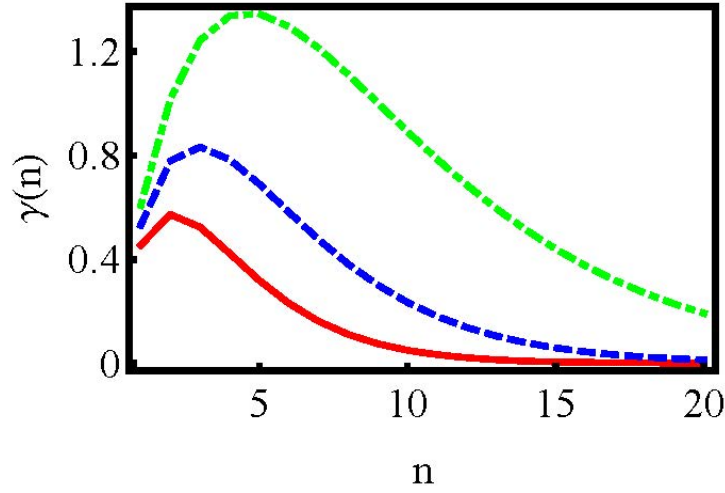

FIG. 7. Dependencies of the interaction constant $|\gamma_n|^2 = \gamma^2 \xi^{2n}(n+1)^2 n^{1/2}(2n+1)^{-1/2}$ on $n$ for different values of $\xi$. $\xi = 0.5$ (solid red line), $\xi = 0.6$ (dashed blue line), and $\xi = 0.7$ (dot-dashed green line).

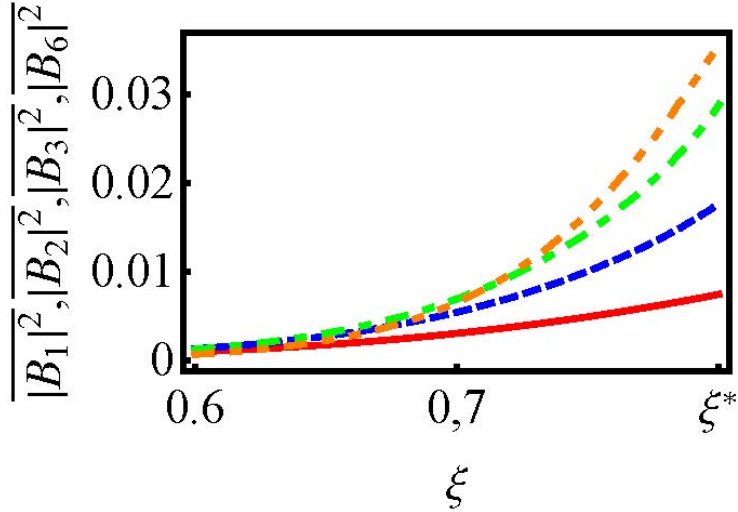

FIG. 8. Dependencies of average probabilities of excitations of the resonator modes $\overline{|B_1(t)|^2}$ (solid line), $\overline{|B_2(t)|^2}$ (dashed line), $\overline{|B_3(t)|^2}$ (dash-dotted line), and $\overline{|B_6(t)|^2}$ (dash-double dotted line) on $\xi$ for $0.5 < \xi < \xi^*$

  The behavior shown in Fig. 8 abruptly changes when the distance between the TLS and NP decreases further (see Fig. 9). At a distance $\xi = \xi^* \approx 0.8$, the probability of the excitation of the $k = 6$-mode peaks out; it follows by peaks of $\overline{|B_3(t)|^2}$ at $\xi \approx 0.87$, then $\overline{|B_2(t)|^2}$ at $\xi \approx 0.9$, and finally by a sharp peak of the excitation probability of the dipole mode at $\xi \approx 0.93$ (Fig. 9).

For further decrease of $\xi$, the mode behavior is reversed again: the peculiarities disappear and $\overline{\left|B_k(t)\right|^2}$ is maximal for the mode that has the strongest interaction with the TLS (see Fig. 10).

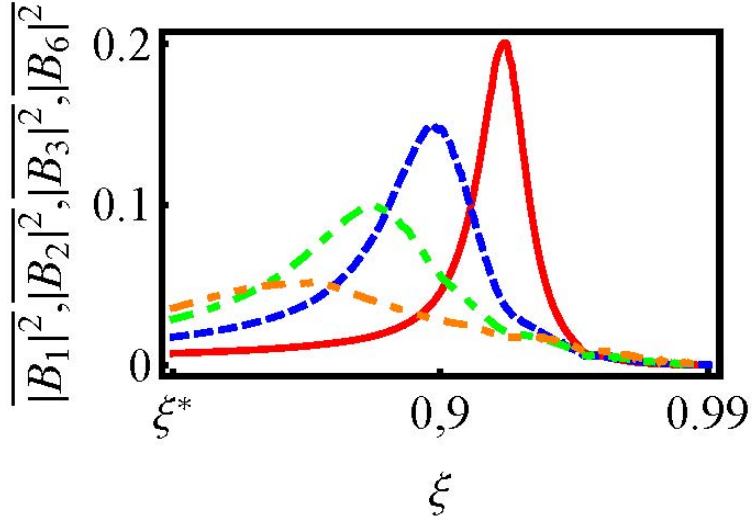

FIG. 9. The same as in Fig. 9 for $0.8 < \xi < 1$ .

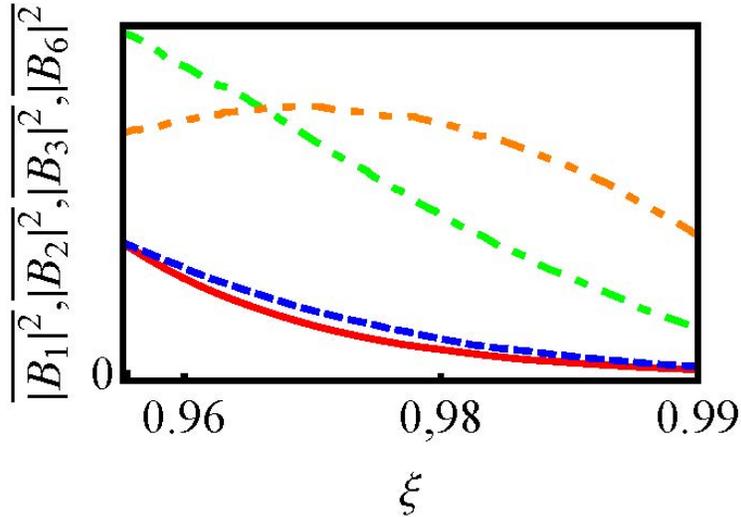

FIG. 10. The same as in Figs. 9 and 10 for $0.94 < \xi < 0.99$ .

This unexpected result can be explained qualitatively. When $\xi$ increases, the interaction with higher multipole modes also increases because the maximum of the interaction constant grows with $\xi$ (see Fig. 7). However, the eigenfrequencies of the multipole modes have the condensation point $\omega_n \to \omega_{pl}/\sqrt{2}$ . Therefore, since we assume that $\omega_{TLS} = \omega_{pl}/\sqrt{2}$ , the interaction of the TLS with higher modes is reduced to the interaction of two oscillators with equal frequencies. The interaction removes the degeneracy. The greater the interaction between oscillators the greater is the difference between eigenfrequencies of the joint system.[14] At some distance between the TLS and NP, the eigenfrequency of the joint system may coincide with one of the lower multipole modes. In this case, the energy is transferred to this particular mode causing near regular Rabi oscillations.

This qualitative argument is supported by analytical estimate. Let us consider Eq. (16) for the Fourier-transform of the probability amplitude $\alpha(q)$ for the population of the TLS excited state at infinite time and $\xi \to 1$. For the Fourier-transform this means that we should evaluate $\alpha(q)$ at zero point:

$$\alpha(q) \sim \frac{1}{q} \frac{1}{1 - \frac{\gamma^2}{q} \sum_{k=1}^{\infty} \frac{\xi^{2k} k^2}{q + 1/k}} \approx \frac{1}{q} \frac{1}{1 - \frac{\gamma^2}{q} \sum_{k=1}^{\infty} \frac{\xi^{2k} k^2}{1/k}} = \frac{1}{q} \frac{1}{1 - \frac{\gamma^2}{q} \text{Li}_{-3}(\xi^2)}$$
$$= \frac{1}{q} \frac{1}{1 - \frac{\gamma^2}{q} \frac{\xi^2 (1 + 4\xi^2 + \xi^4)}{(1-\xi^2)^4}} \approx \frac{1}{q} \frac{1}{1 - \frac{6\gamma^2}{(1-\xi^2)^4 q}}$$
(49)

The poles of the last expression are at

$$q \sim \pm \gamma (1 - \xi^2)^{-2}.$$
(50)

When $\xi \to 1$, the value of $q$ at the pole grows. The Fourier-transform for the $k$-th mode, Eq. (16), shows that the resonance occurs at $q_k = -1/k$. Thus, when the TLS approaches the NP, the effective frequency of oscillations detunes from the frequency $\omega_{TLS} = \omega_{pl}/\sqrt{2}$. When it coincides with $q_k = -1/k$, the resonance with the $k$-th mode occurs. There are no resonances for $q < -1$, because modes with the corresponding frequencies do not exist. This explains the fact that after the dipole resonance is reached, the maximum energy is transformed into the mode with the strongest interaction.

### B. The effect of Joule losses

In this section we take into account the influence of Joule losses on the plasmonic mode excitation described above. Firstly, we recall that for a single mode cavity, dissipation leads to either the monotonic exponential decay of the excited state of the TLS, in the weak coupling regime, or the exponential decay of the Rabi oscillation, in the strong coupling regime. To show this we use Eq. (18) with $|\gamma_k|^2 = \omega_R^2$, $g_k = \gamma_{CM}$. So we can find poles of the integrand of Eq. (18): $q = \omega_R^2 / (q + i\gamma_{CM})$.

In the weak coupling regime, $\omega_R \ll \gamma_{CM}$, at which $q \approx -i\omega_R^2/\gamma_{CM}$, we have the exponential decay with the rate $\omega_R^2/\gamma_{CM}$:

$$A(t) = \exp(-\omega_R^2 t/\gamma_{CM}). \tag{51}$$

In the strong coupling regime, $\omega_R \gg \gamma_{CM}$, at which $q \approx \pm\omega_R - i\gamma_{CM}/2$,

$$A(t) = \exp(-(\gamma_{CM}/2 + i\omega_R)t). \tag{52}$$

In Fig. 11, we show the results of the numerical solution of Eq. (45) in which Joule losses are taken into account. The characteristic relaxation time of the photon radiation of the TLS is much larger than all other times. Indeed, a typical value of the radiation time for an atom and a quantum dot are $\tau_\sigma \sim 10^{-9} s$ and $\tau_\sigma \sim 10^{-11} - 10^{-10} s$, respectively, while the relaxation time due to Joule losses in the NP is of the order of $\tau \sim 10^{-13} - 10^{-12} s$. Thus, we can neglect the photon radiation because the exponential decay due to radiation into the free space is not noticeable on the considered timescales. We assume that $g_k \sim 10^{12} s^{-1}$ for each NP mode. In addition, there is strong loss, $g \sim 10^{13} s^{-1}$, due to radiation in the dipole mode. Loss in the TLS is much smaller than in the NP, therefore it can be neglected. Fig. 11 shows the same average probabilities of excitation of the resonator mode, $\overline{|B_k(t)|^2}$, as shown in Fig. 9.

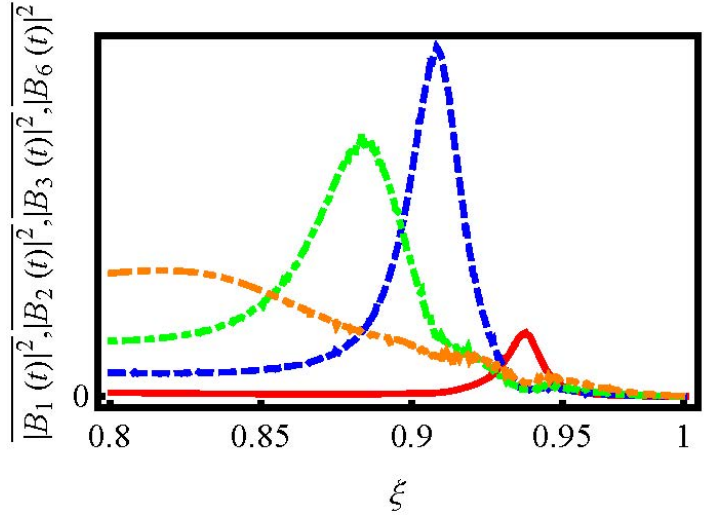

FIG. 11. The same as in Fig. 10 for a lossy system.

As one can expect, when Joule loss is taken into account, the smaller fraction of the TLS energy is transferred to the NP modes. In addition, the dipole mode is excited much less intensively due to the loss for radiation. However, qualitatively the character of the resonances and the mode energy distribution is not changed.

## V. CONCLUSION

In this paper, we consider spontaneous relaxation of a TLS into a condensation point of the plasmon resonances of a spherical NP. We show that when the TLS excitation frequency coincides with the condensation point, the interaction with an infinite number of higher multipole modes plays the main role in the relaxation process. This differs from previous studies in which higher multipoles were considered as corrections.

An infinite countable set of modes plays the role of a reservoir for the TLS relaxation. This reservoir serves as a parallel channel for the exponential decay of TLS excitation and leads to an increase in the relaxation rate. We show, however, that only a fraction of the TLS energy is transferred to the NP and that the rest of the energy oscillates. The envelope of the excited state population of the TLS decays exponentially. The closer the TLS is to the NP, the smaller the oscillating fraction of the energy. At the same time, even though the interaction of the TLS with higher modes is strong, the remaining energy is not transferred into the condensation point but rather into lower modes. The reason for this is the resonant participation of the lower modes in energy transfer. The choice of the resonance frequency is controlled by the distance between the TLS and NP.

The physical picture considered differs from usual radiation quenching of a TLS positioned near a plasmonic structure in which TLS radiation into the free space is suppressed and most of the energy is transferred into plasmons because the probability of plasmon excitation is much higher than radiation of photons. In our system, the radiation is still suppressed, however by changing the distance between the TLS and NP, the energy can be selectively transferred into a desired mode. This result suggests that it may be possible to create a tunable frequency converter in nano-optics.

## ACNOWLEDGEMENTS

This work was partly supported by RFBR Grants Nos. 12-02-01093, 13-02-00407, 13-02-92660, by the Dynasty Foundation, and by the NSF under Grant No. DMR-1312707.